\documentstyle[11pt,epsfig,twocolumn]{article}

\newif\ifamsfonts
\amsfontstrue
\ifamsfonts
\font\twlbbb=msbm10 scaled\magstep1
\font\egtbbb=msbm8
\font\sixbbb=msbm6
\newfam\bbbfam
\textfont\bbbfam=\twlbbb
\scriptfont\bbbfam=\egtbbb
\scriptscriptfont\bbbfam=\sixbbb
\newcommand{\Bbb}[1]{{\fam\bbbfam\relax#1}}
\else
\newcommand{\Bbb}[1]{{\bf#1}}
\fi

\def\Cset{\Bbb{C}}

\def\Nset{\Bbb{N}}

\def\Rset{\Bbb{R}}
\def\Sset{\Bbb{S}}

\newtheorem{thm}{Theorem}[section]
\newtheorem{prop}{Proposition}[section]

\newtheorem{rk}{Remark}[section]

\renewcommand{\theequation}{\thesection.\arabic{equation}}
\makeatletter\@addtoreset{equation}{section}\makeatother

\newcommand{\dn}{\mathop{\rm dn}\nolimits}
\newcommand{\sn}{\mathop{\rm sn}\nolimits}
\newcommand{\cn}{\mathop{\rm cn}\nolimits}

\newcommand{\res}{\mathop{\rm res}\nolimits}

\newcommand{\Or}{\mathop{\rm O}\nolimits}

\newcommand{\id}{\,{\rm d}}

\newcommand{\iu}{\mskip2mu{\rm i}\mskip1mu}

\def\w{\cal}

\def\lgh{\longrightarrow}

\def\pt{\partial}
\def\bd{\begin{displaymath}}
\def\ed{\end{displaymath}}
\def\bqns{\begin{eqnarray*}}
\def\bi{\begin{itemize}}
\def\ei{\end{itemize}}
\def\beq{\begin{quote}}
\def\eeq{\end{quote}}
\def\ben{\begin{enumerate}}
\def\een{\end{enumerate}}
\def\eqns{\end{eqnarray*}}
\def\bq{\begin{equation}}
\def\bqn{\begin{eqnarray}}
\def\eq{\end{equation}}
\def\eqn{\end{eqnarray}}

\def\e{\varepsilon}

\def\g{\gamma}

\renewcommand{\theequation}{\thesection.\arabic{equation}}

\newcounter{saveeqn}
\newcommand{\alpheqn}{\setcounter{saveeqn}{\value{equation}}%
\stepcounter{saveeqn}\setcounter{equation}{0}%
\renewcommand{\theequation}{\mbox{\thesection.\arabic{saveeqn}\alph{equation}}}}
\newcommand{\reseteqn}{\setcounter{equation}{\value{saveeqn}}%
\renewcommand{\theequation}{\thesection.\arabic{equation}}}

\begin{document}

\title{\bf Non-Integrability and Infinite Branching of Solutions of 2DOF
Hamiltonian Systems in Complex Plane of Time
\footnote{To appear in ``Nonlinear Phenomena in Complex Systems", 1999.}}

\author{VASSILIOS M ROTHOS \thanks{Present address: The Nonlinear Centre, 
Department of Applied Mathematics and Theoretical
Physics, University of Cambridge, Silver St., Cambridge
CB3 9EW, U.K}\,\, and\,\, 
TASSOS C BOUNTIS\\ 
Center for Research and Applications of Nonlinear Systems\\
Department of Mathematics, University of
Patras, GR 265 00 Patras, Greece}
\date{}
\maketitle
{\bf  Abstract}
{\footnotesize It has been proved by S.L.Ziglin [1], for a large class of 2-degree-of-freedom (d.o.f)
Hamiltonian systems, that transverse intersections of the invariant
manifolds of saddle fixed points imply infinite branching of 
solutions in the complex time plane and the non--existence of a second
analytic integral of the motion.  Here, we review in detail our recent results,
following a similar approach
to show the existence of infinitely--sheeted solutions for 2 d.o.f.
Hamiltonians which exhibit, upon perturbation, subharmonic bifurcations 
of resonant tori around an elliptic fixed point [2]. Moreover, as shown recently, 
these Hamiltonian systems are non--integrable if their resonant tori
form a dense set. These results can be extended to the case where 
the periodic perturbation is not Hamiltonian.}
\section{Introduction}
\setcounter{equation}{0}

In the last $15$ years, there has been active interest in the study of 
integrability (or absence thereof) of nonlinear dynamical systems based
on the analysis of their singularities in the complex time t--plane [1--6].

Singularity analysis is the study of the behavior of the solutions
of differential equations around their singularities in complex time.
While, any anlytic system of differential equation is locally integrable,
the different local patches do not, in general, fit together globally.
The main idea of the singularity analysis is to obtain global 
information on the integrability of a system through the local
analysis of the solution in the complex plane. Its origins can be found in 
Kowaleskaya's classical work and on the Painlev\'e 's classification of second-order 
ordinary differential equations (see, e.g.[3]). However, for a long
time, Kowaleskaya's
work and Painlev\'e 's theory were consider interesting, if not old
fashioned, masterpieces in the theory of special functions and little
attention was paid to them late 1970's, when it was 
noticed that they were intimately related to the theory of solitons.
The various Painlev\'e tests for ODEs which followed this discovery [3--5] are
based on the formal existence of Laurent expansions for the solutions around
the movable singularities of the solution in the complex plane.

According to this approach one seeks to establish conditions such that all
{\bf movable} (i.e initial condition dependent) singularities of the
solutions of the equations of motion of the system
\bq
{dx\over dt}={\dot x=f(x,t)}\qquad{x=( x_1(t) ,\cdots ,x_n(t) )}
\label{1.1}
\eq
are isolated poles, i.e that (\ref{1.1})
possesses the so--called Painlev\'e property [3--5].
If this is the case, then all solutions of (1.1) are meromorphic
(single--valued and analytic everywhere except at the poles) and the system
is often integrable, in the sense of having global, single--valued
integrals of the motion.

On the other hand, if infinitely multi--valued solutions are found, one
expects that such integrals do not exist and the system is called non--integrable.
The presence of infinitely--sheeted solutions (so--called I.S.S property)
can be deduced either analytically, by showing e.g. that their series
expansions near a singularity contain logarithmic terms, or, numerically, by 
integrating (\ref{1.1}) along contours enclosing one or more singularities in 
$t\in {\Cset}$.

One rigorous approach to the connection between non--integrability and the
I.S.S property was introduced, a number of years ago, by Ziglin [1]. He showed,
under some general assumptions, that 2--d.o.f. Hamiltonian systems of the form
\bq
{\w H}={\w H}_0(x_1, x_2, I) + \e{\w H}_1(x_1, x_2, I, \phi)
\label{1.2}
\eq
with ${\w H}_1$ $2\pi$--periodic in $\phi$, possessing, for $\e=0$,
a closed homoclinic (resp. heteroclinic) orbit, which joins one saddle fixed point
to itself (resp. two saddle fixed points), exhibit, for a large class of
such perturbations and
$0<\vert \e\vert
\ll 1$, infinitely--sheeted solutions.  Ziglin's remarkable discovery was that
he was able to relate directly this I.S.S. property, using Mel'nikov's theory
to the transversal intersections of the stable and unstable manifolds of 
the homo(or hetero)clinic orbit.

Furthermore, Ziglin proved that these perturbed Hamiltonians are non--integrable
in the sense that they cannot have a global, single--valued integral of the 
motion, other than the Hamiltonian itself.

The purpose of this paper is first to review, Ziglin's results on transverse 
intersections of the invariant
manifolds of saddle fixed points and how these imply infinite branching of 
solutions in the complex time plane and the non--existence of a second
analytic integral of the motion. Then we shall prove as reported [2], 
the I.S.S property for Hamiltonian systems
of the form (\ref{1.2}) around {\em an elliptic} fixed point and thus establish a connection
between infinite branching of solutions and the non--integrability  of such  
systems, as long as they exhibit subharmonic bifurcations on {\em a dense}
set of resonant tori [7].

Our approach follows closely that of Ziglin, in that we use one of
Mel'nikov 's theorems to show that I.S.S is a direct consequence of 
subharmonic bifurcations (at $\e\neq 0$) of resonant invariant curves
of the unperturbed ($\e=0$) system. Moreover, we also make the 
crucial assumption that, for $\e=0$, (1.2) possesses only meromorphic
solutions in the complex domain. Unlike Ziglin, however, we do not assume
the presence of a saddle fixed point, whose invariant manifolds
govern the dynamics.

In section 2, we present some useful backgraund information on the structure
and importance of Riemann surfaces, [14].

In section 3, we state Ziglin's theorm on the transversal intersection of
invariant manifolds in the perturbed 2 d.o.f Hamiltonian (\ref{1.2}). For $\e=0$, these
invariant manifolds join ``smoothly'' in a single separatrix, or
homo(hetero)clinic orbit of the completely integrable unperturbed problem,
whose slutions have only poles in complex $t$. For $\e\neq 0$, however, these
manifolds interesct at infinitely many points and infinitely branched
multi--valued solutions appear, as predicted by Ziglin, with e.g. logarithmic
singularities.

In section 4, we state our main theorem, as described in [2], prove that the series expansions
near a singularity contain logatithmic terms using the theory of Abelian
integrals and we illustrate our results on the example of a driven Duffing
oscillator. 

In section 5, we extend and apply our results to the case of non--Hamiltonian
perturbations. Finally, in section 6, we end with some concluding remarks.

\section{Riemann Surfaces}
\setcounter{equation}{0}

A Riemann surface $X$ is a connected two--dimensional topological manifold with
a complex--analytic structure on it.  The latter implies that for each point
$P\in X$ there is a homeomorphism ${\phi}: U\lgh V$ of some
neighborhood $U\ni P$ onto an open set $V\in {\Cset}$, 
and is defined so that any two such
homeomorphisms ${\phi}$, $\hat\phi$ with $U\cap {\hat U}\not=\emptyset$ are
holomorphically compatible, i.e., the mapping ${\phi}\circ {\hat\phi}^{-1}:
{\hat\phi}(U\cap {\hat U})\lgh {\phi}(U\cap {\hat U})$, called a transition
function, is holomorphic.  In what follows, the homeomorphism $\phi$ will be
referred to as a local parameter. Any set ${{\phi}_{_{i}}}$ of holomorphically
compatible local parameters such that the appropriate neighborhoods
${U_{_{i}}}$ cover the entire manifold $X$ is called a complex atlas of the
Riemann surface $X$. The union of the atlases that correspond to the same
complex--analytic structure on the manifold $X$, i.e., to the same Riemann
surface $X$, is again an atlas.  This property is violated if the atlases
making up a union belong to different complex--analytic structures or, which is
equivalent, to different, yet topologically identical Riemann surfaces.

The simplest examples of a Riemann surface are: any open subset of a complex
plane ${\Cset}$, ${\Cset}$ itself or an extended complex plane 
${\Sigma}={\Cset}\cup \{{\infty}\}$. The two canonical examples for the occurence of
multi--valuedness are the $log$ function (for each 
$z\in {\Cset}{\setminus}\{0\}$, $\log(z)$ are the solution of ${\rm
e}^{w}=z$) and the function $z^{{1}/{q}}$ ($q\in {\Nset}$,
$q\geq 2$, solution of $w^{q}=z$).  For these functions, the points $0$ and
$\infty$ are critical, that is, there is no meromorhic continuation around
these points.  However, in any regions 
\[
D_{J}=\Big\{\,z=r{\rm e}^{{\iu}{\theta}}; r\ge 0,
{\theta}\in [a, b],\,a\le b\leq a+2{\pi}\,\Big\}
\]
the function $f$ is
single--valued and analytic.  To define these functions one introduces cuts in
the complex plane, across which the function cannot be continued.  This
approach is not satisfactory since the functions are not continuous on
${\partial}D_{J}$.  The solution to this problem is to extend the domain of
definition, rather than restricting tha values of the function.  It is exactly
this domain, in
which these functions are single--valued, which forms a sheet of the Riemann surface,
in the covering--space of ${\Cset}\setminus \{0\}$. 
For example, each sheet of the surface corresponds to a
particular branch of $\log(z)$.

Meromorhic functions, i.e., non--constant holomorhic mappings $f:X\lgh
{\Sigma}$, constitute a meaningful object of analysis on Riemann surfaces. The
local notation $f(z)=f\circ {\phi}^{-1}(z)$ of a meromorhic function $f$, for
any local parameters $\phi$ is a meromorhic function of the variable $z\in
{\phi}(U)$ in the usual sense. The meromorhic function $f$ takes evey value
$c\in {\Sigma}$ the same finite number of times (with the multiplicity taken
into account). A point $P_{{0}}\in f^{-1}(\infty)$ is said to be pole of
the function $f$. In the neighborhood of any $P_{{0}}\in X$, a meromorhic
function $f$ can be represented a convergent Laurent series:
\bq
\sum_{j=-N}^{\infty}c_{{j}}(z-z_{{0}})^{j},\quad z\equiv {\phi}(P),\quad
z_{{0}}={\phi}(P_{{0}})
\label{2.1}
\eq
where $\phi$ is a local parameter, the number $N> -\infty$ and does not
depend on a specific choice of $\phi$.

In addition to the notion of a function on a Riemann surface, we introduce the
notion of an Abelian differential. An Abelian differential on the Riemann
surface $X$ is a meromorhic 1--form $\omega$, given on $X$. This implies that
we can write $\omega$ locally as $f(z){\id}z$, where $f(z)$ is a meromorphic
function of $z$ in its domain.  For any Abelian differential, the notion of a
pole and that of a zero are precisely defined, along with the notions of
multiplicities and that of a residue:
\bq
{\res}(\omega; P_{{0}})=c_{{-1}},
\quad {\omega}(P)=\sum c_{{j}}(z-z_{{0}})^{j}{\id}z
\label{2.2}
\eq
Abelian differentials are usually divided into three kinds: holomorphic
differentials (first kind), meromorphic differentials with residues equal to
zero at all singular points (second kind), and meromorhic differentials of the
general form (third kind).

Any Abelian differential $\omega$ on the Riemann surface $X$ satisfies the
closure condition
\bq
{\id}{\omega}=0
\label{2.3}
\eq
where ${\id}{\omega}$ 
denotes the total derivative of the 1--form $\omega$.  
Therefore, locally, a primitive function for
the differential $\omega$ always exists and can be defined by the formula
\bq
{\Omega}=\int_{Po}^{P}{\omega}
\label{2.4}
\eq
for any simply--connected domain
on $X$ that involves (in the case of third-kind differentials) no singularities
of the differential $\omega$.  Formula (\ref{2.4}) considered on the whole surface
$X$, defines, in general, a multivalued function called an Abelian integral.
The division of Abelian differentials into the three kinds can be extended to
Abelian integrals. Locally, Abelian integrals of the first kind are holomorhic
functions, Abelian integrals of the second kind are meromorphic functions, and
Abelian integrals of the third kind have logarithmic singularities:
\bq
{\omega}=(\cdots +\frac{c_{{-1}}}{z}+\cdots)\,{\id}z\lgh
{\Omega}=\cdots+c_{{-1}}{\ln}z+\cdots
\label{2.5}
\eq  
We will use expression (\ref{2.5})
to prove that the series expansions of solutions of (\ref{1.2}) near a
singularity contain logarithmic terms, (see Section 4).

\section{Ziglin's theorem on a 2 d.o.f Hamiltonian System}
\setcounter{equation}{0}

In this section, we shall state Ziglin's theorem [1] on the splitting of
separatrices in 2 d.o.f Hamiltonian systems. Consider the Hamiltonian 
\bq
{\w H}={\w H}_0(x_1, x_2, I) + \epsilon{\w H}_1(x_1, x_2, I, \phi)
\label{3.1}
\eq
where $(x, y)$ and $(I, {\phi})$ are canonically conjuagate pairs of
momentum--position and action--angle variables respectively. With Ziglin [1],
we now make the following assumptions about (\ref{3.1}):
\begin{itemize}
\item[{\bf I}] ${\w H}$ is real analytic in some domain of ${\bf x}=(x, y)$,
$\vert I-I_0\vert < \mu$, $\vert \e\vert < \mu$ and 
2$\pi$ - periodic in $\phi$.
\item[{\bf II}] For $\e=0$, $I=I_{{0}}$, (\ref{3.1}) has two hyperbolic fixed points 
${\bf x}_{_{+}}$,  ${\bf x}_{_{-}}$ joined by a doubly asymptotic solution 
${\hat {\bf x}}(t)\lgh {\bf x}_{_{\pm}}$ as $t\lgh\pm\infty$.
\item[{\bf III}] ${\partial}_{I} {\w H}_{_{0}}({\hat {\bf x}}(t), I_{0})\ge
c\ge 0$ for all $t$ and the solution 
\bq
{\hat {\bf z}}(t)=({\hat {\bf x}}(t), I_{0}, {\hat {\phi}}(t)),\quad  
{\hat {\phi}}(t)=\int^{t}\frac{{\partial {\w H}_{{0}}}}{\partial I}
{\id}t^{\prime}
\label{3.2}
\eq
can be analytically continued to the strip 
$${\Pi}=\Big\{ 0\leq {\rm Im}t \leq
2{\pi}/{\lambda}_{{+}}\,\Big\}$$ (where ${\lambda}_{{+}}$ is the positive eigenvalue
of the linearized system about ${\bf x}_{{+}}$) and has no more than a finite
number of singular points in ${\Pi}$.
\item[{\bf IV}] ${\w H}({\bf z}, \e)$ can be analytically continued
for complex ${\bf z}$ and $$\frac{{\partial {\w H}_{{0}}}}{\partial I}({\hat {\bf z}}(t)), 
\frac{{\partial {\w H}_{{1}}}}{\partial {\phi}}({\hat {\bf z}}(t,
{\phi}_{{0}}))$$ 
are single valued in $\Pi$, for all ${\phi}_{{0}}\in {\Rset}$,
where ${\hat {\bf z}}(t, {\phi}_{{0}})$ denotes the solution
${\hat {\bf z}}(t))$ of (3.2) with ${\hat {\phi}}$ replaced by ${\hat {\phi}}
+{\phi}_{{0}}$.
\end{itemize}
\begin{thm}
{\em (Ziglin [1])} Under the above assumptions and if 
${\pt}_{\phi} {\w H}_{1}({\hat {\bf z}}(t,{\phi}_{{0}}))$ 
has nonzero sum of residues (\ref{2.2}) in $\Pi$ (for at least one
${\phi}_{{0}}$), the system (\ref{3.1}) possesses multiple--valued solutions
$I(t)=I_{{0}}+{\e}I_{{1}}+\cdots$ since 
\bq
{\Delta}I_{{1}}={\oint}_{\Gamma}{\dot I}_1\,dt = -{\oint}_{\Gamma}
{\pt}_{\phi}{\w H}_1({\hat {\bf z}}(t, {\phi}_{{0}})){\id}t\neq 0
\label{3.3}
\eq
for some contour $\Gamma\in\Pi$. In fact, since for any given
${\phi}_{{0}}$, going around $\Gamma$ changes the value of $I_{{1}}$ by the
same amount ${\Delta}I_{{1}}$, we conclude that $I(t)$ is infinitely branched
in the complex t--plane, much like a ${\log}t$ function.
\end{thm}

The connection between (\ref{3.3}) above and the splitting of separatrices comes from Ziglin's
proof [1] that (\ref{3.3}) implies that the following integral does not vanish identically:
\bq
J({\phi}_{{0}})=\int_{-\infty}^{\infty}\{ H_0,H_1 \}({\hat {\bf z}}(t),
{\phi}+{\phi}_{{0}}){\id}t\neq 0
\label{3.4}
\eq
where $\{{\cdot} , {\cdot}\}$ denotes the Poisson bracket, and $H_0, H_1$ are related to the
original Hamiltonians by solving (3.1) for a (single--valued) $I$ on a constant
energy surface $-I=H_0(x,y)+{\e}H_1(x, y, {\phi})+\cdots$ .
Ziglin also proves that (\ref{3.3}) implies that the Hamiltonian system (\ref{3.1}) does
not possess a second analytic integral indepedent of ${\w H}$ for any
sufficiently small $\vert \e \vert\neq 0$ [1].

Theorem 3.1 can be generalized to the non--Hamiltonian case of a
periodically driven system
$${\dot x}_{{i}}=f_{{i}}(x_{{1}}, x_{{2}})+
{\e}g_{{i}}(x_{{1}}, x_{{2}}, t),\quad {i=1,2}$$
where $g_{{i}}(x_{{1}}, x_{{2}}, t)=g_{{i}}(x_{{1}}, x_{{2}}, t+T)$, 
$f_{{i}}$, $g_{{i}}$ are analytic in $x_{{1}}$, $x_{{2}}$ and the
unperturbed $(\e=0)$ equations have single--valued solutions and a smooth
separatrix joining two fixed saddle points (see [12]).
The main idea is that if the Mel'nikov integral is not identically zero, one
can always find an analytic function of $x_{{1}}$, $x_{{2}}$ which is
infinitely--sheeted in the complex t--plane. 

Of course, splitting of separatrices does not necessarily mean the
appearance of chaos, since, in a non--Hamiltonian system, invariant manifolds of
saddle points can split, for ${\e}\neq 0$, without intersecting
($J({\phi}_{{0}})\neq 0$ for all ${\phi}_{{0}}$).  Splitting does mean,
however, non--integrability, in the sense of the appearence of I.S.S with
the type of infinite multivaluedness one finds in functions with logarithmic
singularities.

\section{Infinitely Multivalued Solutions and Subharmonic Bifurcations}
\setcounter{equation}{0}

Let $U=D\times(I_0-\mu , I_0+\mu)\times {\Sset}^1$ be the direct product of a domain
$D\subset{\Rset}^2$ with coordinates $x=(x^1,x^2)$ and $I$ be an action variable
in the interval $I_0-\mu<I<I_0+\mu$
with an angular coordinate $\phi$ on the circle ${\Sset}^1$.
Consider Hamiltonian of the form (\ref{3.1}) 
\bq
{{\w H}(z, \e)={\w H}_0(x,I)+
{\e}{\w H}_1(x, I,{\phi})},\qquad z=(x, I, \phi)
\label{4.1}
\eq
which is real--analytic in
$x=(x_1, x_2), I, \phi$, ${\w H}_0(x, I)=F(x)+G(I)$ for
$x\in D$, $\vert I-I_0\vert < \mu$, $\vert \e\vert < \mu$ and is
2$\pi$ - periodic in $\phi$.  For $\epsilon=0$, we have the unperturbed system:
\bqn 
\dot x_1 & = & \pt_{x_2}{\w H}_0(x,I)\cr
\dot x_2 & = & -\pt_{x_1}{\w H}_0(x,I)\cr
\dot \phi & = & \pt_I{\w H}_0(x,I)\cr
\dot I & = & -\pt_{\phi}{\w H}_0(x,I)=0
\label{4.2}
\eqn

We now make the following assumptions :
\begin{itemize}
\item[{\bf A1.}] The system (\ref{4.2}) possesses an elliptic fixed point at (0,0) and a family 
of doubly periodic (in $t\in {\Cset}$) solutions $q^a(t)$, which
rotate about (0,0) in the $(x_1,x_2)$ plane.
Let $a\in[-1,1]$ with  $q^{-1}(t)=(0,0)$ and $q^a(t)\lgh Q(t)$ as $a\lgh 1$,
where $q^1(t)=Q(t)$ is a boundary of $D$.
\item[{\bf A2.}] Region $D$ of the phase plane of the (uncoupled) $F(x)$ system
is filled with periodic orbits $q^{\alpha}(t)$, whose period 
$T_{\alpha}$ varies continuously with respect
to the energy ${\cal H}(z, 0)=h$.  
Each such orbit is a level of $F:F(x)=h_{\alpha}$
provided, for energy $h>h_{\alpha}$, the unperturbed ($\e=0$)
system has a corresponding closed orbit given by 
$G^{-1}(h-h_{\alpha})\equiv I_{\alpha}$.
\item[{\bf A3.}] The Hamiltonian ${\w H}(z,\e)$ can be continued analytically,
but in general, not single--valuedly to a domain in complex $(z,\e)$ - space.
\item[{\bf A4.}] The functions 
\bq
{\pt _I}{\w H}_0(z(t)),\qquad 
{\pt _\phi}{\w H}_1(z(t,\bar\phi))
\label{4.3}
\eq
with $z(t)=(q^a(t), I)$ and $z(t,\bar\phi)=(q^a(t), I, \phi+\bar\phi)$
are single--valued for every $\bar\phi{\in{\Rset}}.$
\end{itemize}

For some $I_0$ and sufficiently small $\vert I-I_0\vert$, $\vert x-q^0\vert$
and $\vert \e\vert$ the equation of the isoenergy surface,
\alpheqn
\bq
{\w H}(x, I, \phi, \e)=h\label{4.4.a}
\eq
has a single--valued solution for I,
\bq
-I=H(x, \phi, \e; h)=H _0(x) + \e H _1(x,\phi; h)+\cdots
\label{4.4.b}
\eq
with 
\bq
{\w H}_0(x, -H_0(x))=h\qquad H_0(x)=G^{-1}(h-F(x))\label{4.4.c}
\eq

\bq
H_1(x, \phi)=
{{\w H}_1(x, -H_0(x), \phi)\over
{\Omega}(H_0(x))}
\label{4.4.d}
\eq
\reseteqn

This allows us to go, on the surface (\ref{4.4.a}) (for sufficiently small $\vert I-I_0\vert$,
$\vert x-q^0\vert$ and  $\vert \e\vert$), from system (\ref{4.1}) to the
reduced system :
\bq
x_1^{\prime}=\pt_{x_2}H(x,\phi,\e)\qquad 
x_2^{\prime}=-\pt_{x_1}H(x,\phi,\e)\label{4.5}
\eq 
(where primes denote differentiation with respect to $\phi$).
We now consider system (\ref{4.5}) in the extended phase space, $V=D^1\times {\Sset}^1$,
which is the direct product of a domain $D^1\subset D$ containing the 
periodic orbit $q^a(t)$ and  ${\Sset}^1$ (with angular coordinate $\phi$) and define
the Poincar\'e map:
\bq
P_{\e}\,:\,(x_1(\phi), x_2(\phi))\lgh (x_1(\phi+2\pi), x_2(\phi+2\pi))
\label{4.6}
\eq
on the solutions of (\ref{4.5}).

Before we proceed to our main theorem, we need to establish an important
proposition which is analogous to Mel'nikov's second theorem on subharmonic
bifurcations, but does not assume the presence of a saddle fixed point 
governing the dynamics of the system.

For fixed $h$, eqs (\ref{4.5}) take the form of a periodically perturbed planar system
since $H_{0}$ depends only on $(x_{1}, x_{2})$ while $H_{1}$ has an explicit
$\phi$--depedence. The unperturbed Hamiltonian (\ref{4.1}) 
has the form ${\w H}_0(x, I)=F(x)+G(I)$
and the period of periodic orbits for this system satisfies the resonance relationship 
\bq
T_{\alpha}=\frac{2\pi m}{n{\Omega}(h-h_{\alpha})}
\label{4.7}
\eq
for relatively prime integer pairs m, n.  In terms of the ``new time'' $\phi$,
however, the period ${\hat T}_{\alpha}$ of the $H$ system is given by 
$${\hat T}_{\alpha}=\frac{2\pi m}{n}$$
\begin{prop}
{\em (see [2])}
Let $\{H_0,H_1\}$ denote the Poisson bracket of $H_0, H_1$ with respect to $x_1,
x_2$ and consider a periodic orbit of the unperturbed system $q^{\alpha}(\phi)$
with period ${\hat T}_{\alpha}=2\pi m/n$ on the resonant closed curve ${\w H}_0=h_{\alpha}$
of the unperturbed Poincar\'e map $P_0$, cf. (\ref{4.6}). 
Fix $h>0$, m, n integer relatively prime and choose $\epsilon$ sufficiently small.
Then, if the subharmonic Mel'nikov function
\bq
M^{m/n}(\bar\phi)=\int _{0}^{2\pi m}\{ H_0,H_1 \}(q^a(\phi
-\bar\phi), \phi; h){\id}{\phi}
\label{4.8}
\eq
has j simple zeros as a function of $\bar\phi\in [0, 2{\pi}m/n)$,
the resonant torus given by $(q^{\alpha}(\phi-\bar\phi), \phi)$  breaks into $2k=j/m$
distinct $2{\pi}m$--periodic orbits, and there are no other $2{\pi}m$ periodic
orbits in its neighborhood.
\end{prop} 

We now proceed to our main theorem [2]: 

\begin{thm}
Consider a real analytic 2 d.o.f Hamiltonian system (\ref{4.1}), which can be reduced
to the form (\ref{4.5}), by virtue of assumptions ${\bf A1}$-${\bf A4}$, and let $q^{\alpha}(t)$
be a periodic solution of the unperturbed problem, $\epsilon=0$, with ${\w H}_0=
h_{\alpha}$.  Assume, furthermore, that $q^{\alpha}(t)$ is a meromorphic function
of $t\in {\Cset}$, doubly periodic with real and imaginary periods $T_{\alpha}$,
$iT^{\prime}_{\alpha}$ respectively and can be analytically continued in the strip
$$L=\Big\{\,t\in {\Cset}\,:\,0< {\rm Im}t < T^{\prime}_{\alpha}\, \Big\}$$
Finally, suppose that ${\pt}_{\phi}{\w H}_1(q^{\alpha}(t), \bar\phi)$
(where by $\bar\phi$ we abbreviate $\phi+\bar\phi$) has a finite number of
poles, inside a closed contour ${\Gamma}\subset {\Pi}$, where $\Pi$ is the period
parallelogram
$$\Pi=\Big\{\,t\in {\Cset}\,:\,0 < {\rm Re}t < T_{\alpha},\quad 0<
{\rm Im}t < T^{\prime}_{\alpha}\,\Big\}$$
Then, if for some $\bar\phi$, the sum of the residues at these poles
\bq
S=\sum _{t_j\in{\Pi}}{\res}\{{\pt}_{\phi}{\w H}_1(q^{\alpha}(t_j),
\bar\phi)\}\neq 0
\label{4.9}
\eq
and $\e$ is small enough:

\begin{itemize}
\item[(i)]\,The ${\w H}_0=h_{\alpha}$ invariant curve undergoes a subharmonic
bifurcation.
\item[(ii)]\,The system possesses ${\Or}(\e)$ solutions, which are
infinitely--sheeted in $t\in {\Cset}$.
\end{itemize}
\end{thm}

\begin{rk}
{\em The second result, (ii), follows immediately from Hamilton's equations and 
assumption (\ref{4.9}): Evaluating the solution $I(t)=I_0+\e I_1(t)+\cdots$
along a closed contour ${\Gamma}\subset {\Pi}$ we have to ${\Or}(\e)$:
\bqns
{\Delta}I_1 & = & {\oint}_{\Gamma}{\dot I}_1{\id}t = -{\oint}_{\Gamma}
{\pt}_{\phi}{\w H}_1(q^{\alpha}(t), \bar\phi){\id}t\cr\cr
& = & -2{\pi}{\iu}S\neq 0
\eqns
Hence, $I(t)$ increases, after every circuit around $\Gamma$, by a fixed amount 
${\Delta}I(\e)$ such that 
$$\displaystyle\lim_{\e\to 0}{{\Delta}I(\e)\over {\e}}\neq 0$$
and the system possesses the I.S.S property.
The quantity ${\Delta}I_1$ is an Abelian integral on a Riemann surface
${\Sigma}$ (see section 2) and we note that the 1--form $\omega$ (third kind),
given by ${\omega}={\pt}_{\phi}{\w H}_1(q^{\alpha}(t), \bar\phi)dt$, has a
finite number of poles and the Abelian integral ${\Delta}I_1$ a logarithmic
singularities. Then $I(t)$ is infinitely branched in the complex plane of time
much like a ${\log}t$ function.}
\end{rk}

\begin{rk}
{\em Finally, we can show, using the results of [7], that under 
the above conditions, the reduced system (\ref{4.5}) is non--integrable.
This is done by noting that due to (\ref{4.7}), our system exhibits subharmonic
bifurcations on a dense $(n, m)$ set of invariant tori and hence according to
the theorems given in [7], it cannot possess a second analytic, single--valued
integral of the motion.}
\end{rk}

Let us now illustrate Theorem 4.1 on the example of a Duffing
oscillator, which does not have a fixed saddle point and is perturbed by a periodic
function, which preserves the Hamiltonian formulation of the system.
More specifically, we consider the system of o.d.es
\bq
{\dot x}_1=x_2\qquad{\dot x}_2=-x_1-{x_1}^3 + {\e}\cos{\omega}t
\label{4.10}
\eq
Note that we can always introduce an angle variable $\phi={\omega}t$ and a conjugate
action variable I, such that the Hamiltonian of this system can be written
in the form (4.4b), i.e
\bq
H=H_0 + {\e}H_1={{x_2}^2\over 2} + {{x_1}^2\over 2} +
{{x_1}^4\over 4}-{\e}x_1\cos{\phi} + I{\omega}
\label{4.11}
\eq                       
It is easy to verify that the unperturbed system $({\e}=0)$ 
possesses a family of periodic orbits around the 
elliptic fixed point $(0,0)$, given in terms of the Jacobi elliptic functions
[10, 11]
\bq
q^k(t)=A\,\big(\,\cn(\lambda t,k), -{\lambda}\sn(\lambda t,k)
\dn(\lambda t,k)\,\big)
\label{4.12}
\eq
with $A^2=2{\lambda}^2k^2$ and ${\lambda}^2(2k^2-1)=-1.$
Clearly,
this family of periodic orbits can be analytically continued in the strip :  
$$ L=\Big\{\,t\in{\Cset}\,:\, 0 \leq {\rm Im}t\leq 2K^{\prime}\,\Big\}$$
where $K$, $K^{\prime}$ are the elliptic integrals of first and second kind.
Since within any closed contour in $L$, the number of poles is finite, this satisfies
the corresponding 
requirement of Theorem 4.1, cf.(\ref{4.9}).

As is well--known [8, 9] subharmonic bifurcations occur in this problem 
when the Mel'nikov integral
\bq
M^{m/n}(\bar\phi)=\int _{0}^{nT_{\alpha}} \{ H_0,H_1 \}(q^k(t) ,t +
\bar\phi){\id}t
\label{4.13}
\eq
has simple zeros, with $nT_a=2{\pi}m$.
We now compute the integral (\ref{4.13}) by the method of residues and find that
for $n=1$
\[
M^{m/1}(t_0)=\frac{(-1)^m{\pi}^2m}{K\sqrt{2-4k^2}}
{\exp}\big[-{{\pi}mK^{\prime}\over 2K}\big]\sin{\omega}t_0
\]
and for $n\neq 1$ 
\bq
M^{m/n}(t_0)=\sqrt2{\pi}{\omega}(2\sin{\omega}t_0-\cos{\omega}t_0)
\sinh{K^{\prime}{\beta}}
\label{4.14}
\eq
and
\[
S
=\sum _{t_j\in{\Pi}} {\res}(\cn({\lambda}t_j, k)\sin{\phi})\neq 0
\]
where $\Pi$ is the period parallelogram of the $\cn$ function. Thus, according 
to Theorem 4.1,
${\Delta}I_1=-2{\pi}{\iu}S\neq 0$ and the infinitely--sheeted property
of the solution $I(t)=I_0+\e I_1(t)+\cdots$, to ${\Or}(\e)$,
has been established. We recall that the existence of such infinitely--sheeted
solutions for a Duffing oscillator similar to (\ref{4.10}) has been explicitly
demonstrated elsewhere [12, 13], in the form of so--called
psi series expansions, involving logarithmic singularities.

\section{The case of non--Hamiltonian perturbations}
\setcounter{equation}{0}

Consider a Hamiltonian system of the form (\ref{4.1}), (\ref{4.2}) perturbed by
dissipative terms as follows:
\bqn
\dot x_1 & = & \pt_{x_2}F(x)+\e\pt_{x_2}{\w H}_1+\e{\gamma}_1f_1\cr 
\dot x_2 & = & -\pt_{x_1}F(x)-\e\pt_{x_1}{\w H}_1+\e{\gamma}_2f_2\cr 
\dot \phi & = & \Omega(I)+\e\pt_I{\w H}_1+\e{\delta}_1g_1\cr
\dot I & = & -\e\{\pt_{\phi}{\w H}_1-{\delta}_2g_2\}
\label{5.1}
\eqn
where $F, G$ and ${\w H}_1$ are as defined in section 4 and $f_i$, $g_i$ are 
analytic functions
of $(x_1, x_2, I)$ and $2\pi$--periodic in $\phi$.  Furthermore we assume that the
$f_i$, $g_i$  are such that the above perturbation is of non--Hamiltonian
character.

In this case, the function 
${\w H}={\w H}_0(x_1, x_2, I)+\e{\w H}_1(x_1, x_2, I, \phi)$,
with ${\w H}_0=F(x_1, x_2)+G(I)$, is no longer
conserved, as it satisfies
\bq
{\dot {\w H}}=\e\bigl\lgroup {\gamma}_1f_1\pt_{x_1}{\w H}
+{\gamma}_2f_2\pt_{x_2}{\w H}+{\delta}_2g_2\pt_I{\w H}
+{\delta}_1g_1\pt_{\phi}{\w H}\bigr\rgroup
\label{5.2}
\eq
or to order $\e$,
\bq
{\dot {\w H}}=\e h(x_1, x_2, I)+{\Or}({\e}^2)
\label{5.3}
\eq
with $h(x_1, x_2, I)={\gamma}_1f_1\pt_{x_1}F+{\gamma}_2f_2\pt_{x_2}F+
{\delta}_2g_2{\Omega}(I)$.

We now apply the implicit function theorem and solve the equation ${\w H}(x, I,
\phi)={\w H}$
for I 
\[
I=H_0(x_1, x_2; {\w H})+\e H_1(x_1 ,x_2, \phi;{\w H})+\cdots
\]
whence, after some calculations, the reduced system has the form:
\bqns
{\dot x}_1 & = & -\frac{\pt H_0}{\pt x_2}-\e\Big[\frac{H_1}
{\pt x_2}-\frac{{\gamma}_1f_1}{\Omega}
-\frac{\pt H_0}{\pt x_2}\frac{{\delta}_1g_1}{\Omega}\Big]+
{\Or}({\e}^2)\cr\cr
{\dot x}_2 & = & \frac{\pt H_0}{\pt x_1}+\e \Big[
\frac{\pt H_1}{\pt x_1} 
-\frac{{\gamma}_2f_2}{\Omega}
-\frac{\pt H_0}{\pt x_1}\frac{{\delta}_1g_1}{\Omega}\Big]+
{\Or}({\e}^2)
\eqns
\bq
{\dot {\w H}}=\e \Omega\Big[ {\delta}_2g_2-\frac{\pt
H_0}{\pt x_1} 
{\gamma}_1f_1-\frac{\pt H_0}{\pt x_2}{\gamma}_2f_2\Big]+{\Or}({\e}^2)
\label{5.4}
\eq

Let $d=({\gamma}_1, {\gamma}_2, {\delta}_1, {\delta}_2)$ denote the dissipation
coefficients in our system.  Also, let $P_{{\epsilon},d}$ denote the Poincar\'e
map associated with the system (\ref{5.4}). As before, we consider a
function ${\w H}={\w H}(x_1, x_2, I, \phi, \e)$ which can be 
continued analytically, but not, in general, single--valuedly
to a domain in complex space, while the function
\bq
R(q^{\alpha}(t), I, \phi+\bar\phi)=\{{{\pt {\w H}_1}\over {\pt \phi}}-{\delta}_2g_2\}(
q^{\alpha}(t), I, \phi+\bar\phi)
\label{5.5}
\eq
is single--valued for every $\bar\phi$.

We now extend the main theorem of section 4 to (\ref{5.1}), assuming that the function
(\ref{5.5}) has a finite number of poles
inside a closed contour ${\Gamma}\subset {\Pi}$, such that
\bq
S=\sum_{t_j\in{\Pi}} {\res}\bigl\lgroup {{\pt {\w H}_1}\over {\pt \phi}}-{\delta}_2g_2
\bigr\rgroup\neq 0
\label{5.6}
\eq

Thus, we conclude that the non--Hamiltonian system (5.1) possesses infinitely sheeted solutions
to ${\Or}(\e)$, of the form $I(t)=I_0+\epsilon I_1(t)+\cdots$, since
\bqn
{\Delta}I_1 & = & {\oint}_{\Gamma}{\dot I}_1{\id}t=-{\oint}_{\Gamma}\,
\Big\{\,{\pt}_{\phi}{\w H}_1(q^{\alpha}(t), \bar\phi)\cr\cr & - & {\delta}_2g_2(q^{\alpha}
(t), \bar\phi)\,\Big\}{\id}t = -2{\pi}{\iu}S\neq 0
\label{5.7}
\eqn

Finally, it is easy to verify, using the ideas of Theorem 4.1, that
condition (\ref{5.6}) also implies that the Mel'nikov function of system
(\ref{5.4})
\bq
M^{m/n}_d(\bar\phi)=M^{m/n}(\bar\phi)+{\hat M}^{m/n}_{\g}(\bar\phi)
\label{5.8}
\eq
with 
\bq
{\hat M}^{m/n}_{\g}(\bar\phi)=
{1\over {\Omega}}\int_0^{2{\pi}m}\Big[{{\pt H_0}\over {\pt x_2}}{\gamma}_2
f_2-{{\pt H_0}\over {\pt x_1}}{\gamma}_1f_1\Big]
{\id}\phi
\label{5.8a}
\eq
cf. (\ref{5.8}) is not identically zero, for
general perturbations $f_i, g_i$. Hence, if (\ref{5.8})
turns out to have simple zeros, the perturbed system will exhibit subharmonic
bifurcations as expected.

Applying this approach to non--Hamiltonian perturbations (\ref{5.1}) of the example
of section 4, we find that the formula (\ref{5.8a}) yields:
\bq
{\hat M}^{m/n}_{{\gamma}_{2}}=-{\lambda}{\gamma}_{2}\int_{0}^{4nK}
{\sn}^2(u,k){\dn}^2(u,k){\id}u
\label{5.9}
\eq
where we have taken $f_{1}=0$, $f_{2}=x_{2}$, $g_{1}=g_{2}=0$ cf. (\ref{4.12}).
Performing the integrations in (\ref{5.9}) we finally find 
\bq
{\hat M}^{m/n}_{{\gamma}_{2}}={\gamma}_{2}
\frac{4}{3k^2{\sqrt {1-2k^2}}}[E(k)(1-2k^2)-{k^{\prime}}^2K(k)]
\label{5.10}
\eq
When added to (\ref{4.14}), the constant 
${\hat M}^{m/n}_{{\gamma}_{2}}$ for $n=1$, obtained above for suitable choices of ${\gamma}_{2}$
\bq
{\gamma}_{2}\le R^{m}(\omega)=\frac{{\hat M}^{m/1}_{{\gamma}_{2}}}
{(-1)^{m}{\gamma}_{2}{\sqrt 2}{\pi}{\omega}\exp[-\frac{m{\pi}K^{\prime}}{2K}]}
\label{5.11}
\eq
can prevent the total $M^{m/1}_{d}({\bar\phi}, {\gamma}_{2})$ of (\ref{5.8}) from
having simple zeros, in $\bar\phi={\omega}t_{0}$, thus eliminate the
occurrence of subharmonic bifurcations in this example and the unperturbed
periodic orbit (\ref{4.12}) satisfying the resonance relation
$$4K(k){\sqrt {1-2k^2}}=\frac{2{\pi}m}{\omega}$$

\section{Conclusions}

In this paper, we have shown that 2 d.o.f Hamiltonians which exhibit,
upon perturbation, subharmonic bifurcations of their resonant invariant curves
around an elliptic fixed point possess solutions which are
infinitely--sheeted in the complex domain. Such systems are known to be
non--integrable when these bifurcating tori
form a dense set.

Our analysis follows Ziglin's approach to similar systems, with
the important difference that, in this case, the unperturbed Hamiltonian possesses
a homoclinic orbit. In our systems, we do not require the existence of such an
orbit and make  crucial use of Mel'nikov's subharmonic function.
We have applied our results to a conservative Duffing oscillator,
showing that this classical example exhibits infinitely--sheeted
solutions near an elliptic fixed point.

Finally, we have shown that our results can be extended to
the case of non--Hamiltonian perturbations,
and have illustrated the analysis by applying to a class of non--Hamiltonian
perturbation of Duffing's oscillator.

\section*{Acknowledgements}

One of us (V.M.R) acknowledges the State Scholarships Foundation (S.S.F)
of Greece. This research was partially supported by a P.EN.ED-G.S.R.T grant 
of the Greek Ministry of Development and the Karatheodori Research 
Program of the University of Patras.

\end{document}